\documentclass[aps,prl,twocolumn,superscriptaddress]{revtex4}

\usepackage{graphicx}
\usepackage{amsmath} 
\usepackage{amssymb}
\usepackage{array}

\newcommand{\smb}{SmB$_{6}$}

\newcommand{\eg}{{\it e.g.}}
\newcommand{\ie}{{\it i.e.}}

\begin{document}

\title{One-dimensional edge state transport in a topological Kondo insulator}

\author{Yasuyuki~Nakajima}
\author{Paul~Syers}
\author{Xiangfeng~Wang}
\author{Renxiong~Wang}
 \affiliation{Center for Nanophysics and Advanced Materials, Department of Physics, University of Maryland, College Park, MD 20742}
\author{Johnpierre~Paglione}
 \email{paglione@umd.edu}
 \affiliation{Center for Nanophysics and Advanced Materials, Department of Physics, University of Maryland, College Park, MD 20742}

\date{\today}

\begin{abstract}

Topological insulators, with metallic boundary states protected against time-reversal-invariant perturbations \cite{hasan10}, are a promising avenue for realizing exotic quantum states of matter including various excitations of collective modes predicted in particle physics, such as Majorana fermions \cite{wilcz09} and axions \cite{wilcz87}. 
According to theoretical predictions \cite{dzero10}, a topological insulating state can emerge from not only a weakly interacting system with strong spin-orbit coupling, but also in insulators driven by strong electron correlations. The Kondo insulator compound \smb\ is an ideal candidate for realizing this exotic state of matter, with hybridization between itinerant conduction electrons and localized $f$-electrons driving an insulating gap and metallic surface states at low temperatures \cite{takim11}. 
Here we exploit the existence of surface ferromagnetism in \smb\ to investigate the topological nature of metallic surface states by studying magnetotransport properties at very low temperatures. 
We find evidence of one-dimensional surface transport with a quantized conductance value of $e^2/h$ originating from the chiral edge channels of ferromagnetic domain walls, providing strong evidence that topologically non-trivial surface states exist in \smb.

\end{abstract}


\maketitle

First reported over 40 years ago \cite{menth69}, \smb\ is a prototypical Kondo insulator with hybridization between itinerant conduction electrons and localized $f$ electrons causing an energy gap to open at the Fermi energy and an insulating state to appear on cooling temperatures below the onset of the hybridization gap. 
Recent theoretical calculations \cite{dzero10} have suggested \smb\ as a candidate for realizing the first strongly correlated version of a three-dimensional strong topological insulator (TI) \cite{ando13}.
If true, this could not only solve a long-standing puzzle involving the saturation of electrical resistivity in \smb\ at low temperatures \cite{menth69}, but also provide the first case of a truly insulating stoichiometric TI material.

Several recent experimental studies \cite{wolga13,kim13a,zhang13,xu13,neupa13,jiang13,frant13,zhu13,kim14,thoma13} of \smb\ have provided circumstantial evidence for the existence of topologically non-trivial metallic surface states.  However, the existence of polarity-driven metallic surface states \cite{zhu13} and lack of direct evidence of the chiral nature of surface conduction has brought into question the experimental evidence for TI surface states.
Here we show a signature of non-trivial topological metallic surface states in \smb\ as revealed by one-dimensional conduction along domain wall edges in a naturally occurring surface ferromagnetic state. Together with a suppression of weak antilocalization by spin-flip scattering, anomalous Hall effect (AHE), a hysteretic irreversibility in magnetoresistance and an unusual enhanced domain wall 
conduction provide evidence of long range magnetic order that gaps the Dirac spectrum of the topological surface states and relegates conduction to chiral edge channels.
 
Recent transport experiments \cite{wolga13,kim13a,kim14,thoma13} have proven the existence of metallic conduction at the surface of \smb\ crystals at temperatures much below the opening of the hybridization gap. In this limit, the surface conductance dominates that of the insulating bulk of the crystal, as shown by nonlocal transport \cite{wolga13} and sample thickness dependence studies \cite{syers15}. 
The overall magnetoresistance (MR) in \smb\ is negative at low temperatures and varies quadratically with field, which can be attributed to the reduction of the Kondo energy gap by magnetic field and the liberation of bulk charge carriers \cite{coole95a}. In Fig.~1, we present measurements of the four-wire MR for the slab-shape polished sample with electrical contacts applied by silver paint on one side (see Supplementary Materials (SM)). The MR is measured in perpendicular field orientation $(H_{\perp}\equiv H \parallel [001], I \parallel [100])$ and at temperatures between 20~mK and 1~K, where the surface conduction dominates that of the bulk. 

The MR measurements obtained while applying increasing (up-sweep) field, or $H_{up}$ (Fig. 1a), are qualitatively similar to those taken upon decreasing (down-sweep) field, or $H_{dn}$ (Fig. 1b), but with notable differences. 
For instance, below 500~mK an oscillatory behavior in the MR is visible in the up-sweep data reminiscent of Shubnikov-de Haas oscillations, while it is nearly absent in the down-sweep MR data as shown in Fig.~1b. 
Furthermore, upon close inspection of the down-sweep MR data, abrupt transition-like features are apparent at low $H_{dn}$ fields that are completely absent in the up-sweep data. Below 200~mK, the magnetoresistance abruptly drops through transition-like steps at low $H_{dn}$ fields, as shown in the inset of Fig.~1b. To compare directly, Fig.~2a presents 100~mK MR data obtained by systematically sweeping through the full ``four-quadrant'' range, revealing a stark contrast between up- and down-sweep MR. This takes the form of a hysteretic loop that does not depend on the sign of the field, but only on the sweep direction. The hysteretic loop appears to close at a turning field of $\sim 10$~T, with no difference in $H_{up}$ or $H_{dn}$ MR above that field, and vanishes if the turning field is less than 4~T (see SM).

Shown in Fig.~2b, the MR hysteresis also depends on magnetic field orientation with respect to the sample. When the field is oriented parallel to the surface with electrical contacts $(H_{\parallel}\equiv H \parallel [010], I \parallel [100])$, the difference in up- and down-sweep MR becomes vanishingly small in magnitude. Since it is very unlikely that a bulk-origin anomaly would break the cubic symmetry of the crystal, it is clear that this anomalous MR hysteresis stems from surface conduction.

The observation of weak antilocalization (WAL) confirms this picture. In 2D conductors, weak localization appears as a quantum correction to classical magnetoresistance caused by the constructive or destructive interference between time-reversed quasiparticle paths. The presence of strong spin-orbit coupling or a $\pi$ Berry's phase associated with the helical states of a topological insulator \cite{fu07a} changes the sign of the correction and gives rise to the signature WAL of enhancement of conductance, which is suppressed by a time reversal symmetry-breaking perturbation. An applied magnetic field thus destroys the WAL effect, as described by the Hikami-Larkin-Nagaoka (HLN) equation \cite{hikam80}, 
\begin{equation}
\Delta G_s=-\alpha e^{2}/2\pi^{2}\hbar[\ln(H_{0}/H)-\Psi (H_{0}/H+1/2)], 
\end{equation}
where $\Delta G_s$ is a correction of sheet conductance, $\alpha$ is a WAL parameter, $\Psi (x)$ is the digamma function, $H_{0}=\hbar/4eL_{\phi}^{2}$ and $L_{\phi}$ is the dephasing length.

The WAL effect is normally only sensitive to the perpendicular component of the magnetic field, as it is an orbital effect. Surprisingly, we observe WAL at low temperatures in both $H_{\perp}$ and $H_{\parallel}$ field orientations, but with greatly differing correction amplitudes as shown in Figs.~2c and d. The observation of WAL in $H_{\parallel}$ field orientation most likely originates from a finite width of the surface conducting state wave function that penetrates into the bulk with a characteristic length $\lambda$. In weak correlated topological insulators, such as Bi$_2$Se$_3$, the penetration depth $\lambda$ is negligibly small compared with the dephasing length \cite{zhang10a}. By contrast, in {\smb}, the penetration depth $\lambda \sim \hbar v_F/\Delta$, where $\Delta$ is the bulk gap size, can be longer due to the small gap size of $\Delta\sim$40~K \cite{coole95,wolga13}. Indeed, the long penetration depth in {\smb} is consistent with an extremely high surface carrier concentration of $n_{2D}\sim 10^{14}$ cm$^{-2}$ observed in gating studies \cite{syers15}, in fact much higher than that of known confined electron systems \cite{wolga13}. A finite value of $\lambda$ allows orbital motions of electrons even in $H_{\parallel}$ field orientation, leading to a WAL correction in the sheet conductance given by,
\begin{equation}
\Delta G_s=-\alpha e^{2}/2\pi^{2}\hbar[\ln(1+(H/H_{\parallel})^2)], 
\end{equation}
where $H_{\parallel}=\hbar/\sqrt{2}eL_{\phi}\lambda$ \cite{tkach11}.

As shown in Fig.~2c, we fit the 20~mK low-field sheet conductance to the HLN formula and extract the dephasing length $L_{\phi}$ and the $\alpha$ parameter, expected to be $\alpha =1/2$ for one independent conduction channel. While the extracted $L_{\phi}$ is comparable to previous results \cite{thoma13}, the obtained $\alpha_{\perp}=$ 0.17 in $H_{\perp}$ field orientation is much smaller than the expected value of $\alpha = 2\times 1/2=1$ from top and bottom conduction channels per Dirac cone \cite{thoma13}. Extracting parameters for $H_{\parallel}$ field orientation using Eq.~2 (Fig.~2d) yields a penetration depth of $\lambda=$142~nm (larger than in weakly correlated TIs and comparable to the dephasing length) and $\alpha_{\parallel}=$0.29, which is a small value but still much larger than $\alpha_{\perp}$.

We attribute the strongly suppressed values for both $\alpha_{\perp}$ and $\alpha_{\parallel}$ to the presence of spin-flip scattering (omitted in Eqs.~1 and 2). Spin-flip scattering reduces WAL due to destructive interference, leading to $\alpha = 0$ in the extreme limit where spin-flip scattering is much stronger than spin-orbit scattering \cite{hikam80}. In \smb, spin-flip scattering centers likely nucleate from unscreened $f$-electron Sm moments (so-called ``Kondo holes'' \cite{hamid11}) that have been proposed to explain logarithmic corrections to surface conductance at low temperatures \cite{thoma13}, or possibly Sm$^{3+}$ moments that originate from a surface oxide layer confirmed by X-ray photoelectron spectroscopy \cite{phela14}. 

Whatever their origin, these moments play a similar role to that of magnetic impurities in forming a ferromagnetic (FM) state on the surface of a TI system \cite{liu09a}. In the presence of conducting TI surface states, magnetic order can be stabilized via RKKY interactions, and is guaranteed to be of the FM type if the chemical potential is close to the Dirac point due to a small Fermi wave number \cite{liu09a}. In \smb, surface FM order, characterized by a Curie temperature of $T_C \simeq 600$~mK as shown by the onset of hysteresis in Fig.~3a, naturally explains both the hysteresis observed in the MR and the unexpected anisotropy in the WAL effect.
Moreover, it explains our observation of its hallmark signature, the anomalous Hall effect \cite{nagao10}. As shown in Fig.~3b, the Hall resistance is completely linear above $T_C$ (\ie, 1~K) and its sign is negative, consistent with previous reports \cite{allen79,coole95,syers15}. At low temperatures below $T_C$, a kink is clearly discernable in the raw data precisely near 8~T, the field at which the hysteretic loop in MR closes. Plotted with a linear background subtracted, the difference $\Delta R_{yx}=R_{yx}- AH$ (where $A$ is a linear coefficient obtained from fitting $R_{yx}$ below 5 T) is shown in Fig.~3c to exhibit an abrupt onset, rather than a curvature, suggesting that the observed Hall resistance has the AHE term $R_{yx}^A$ associated with FM domain alignment. Finally, the observation of domain wall dynamics, as indicated by the presence of a strongly asymmetric relaxation in MR between $H_{up}$ and $H_{dn}$ (see SM), confirms without a doubt the presence of FM domains.

Together with hysteresis and AHE observations, these observations are most readily explained by the presence of surface-based FM order in \smb.
The associated hysteretic MR loop is, however, quite different from the conventional butterfly shape observed in common FM materials. First, it is not centered around zero magnetic field. This can indeed occur in certain situations (\eg, exchange bias (Fig.~3d) \cite{giri11,mccor09} or `negative' hysteresis \cite{esho76}), but it is not consistent with the usual overshoot that is necessary to overcome a coercive field. Second, the increased scattering observed in \smb\ upon decreasing field (\ie, $R(H_{dn}) > R(H_{up})$) is opposite to that usually observed in a ferromagnet, where scattering associated with domain walls is typically enhanced on magnetization reversal.
Rather, there is an enhanced conductance in \smb\ upon up-sweep that is diminished upon reaching the turning field and returning to low fields.
Equivalent behavior has been observed in ferromagnetic Mn-doped Bi$_{2}$(Te,Se)$_{3}$ thin films tuned by ionic liquid gating techniques \cite{check12}. In this magnetic TI system, a reversal of the usual hysteresis butterfly shape occurs upon gating the system into the bulk gap regime, where the TI chiral conducting modes trapped by domain walls result in an anomalous Hall conductance associated with a quantum Hall droplet \cite{hasan10}. In this picture, the domain-wall conductance is enhanced during reversal of the magnetization because the number of domain walls increases; at the coercive field, where the number of the domain walls is a maximum, the conductance exhibits a maximum. 

In \smb, this is readily shown by plotting the difference in conductance, $\Delta G$=$G_{up}$--$G_{dn}$ (see Fig.~3e), where $G_{up}~(G_{dn})$ is the magnetoconductance for $H_{up}$~($H_{dn}$). With decreasing temperature, $\Delta G$ is gradually enhanced and the peak position shifts to higher field. The temperature dependence of this characteristic field $H^{\ast}$ follows a mean-field-like order parameter dependence that terminates at the Curie temperature $T_{C}$=600~mK, as shown in Fig.~3c inset. We therefore interpret $H^{\ast}$ as a coercive field, in terms of the enhancement of the conductance. 
Below 100 mK, the peak conductance at $H^{\ast}$ reaches a value of $\sim e^{2}/h$. This value is also observed in several samples with very different values of measured resistance and sample dimensions (see SM), indicating that the observed conductance of $e^{2}/h$ is not coincidental and possibly quantized.
To explain the anomalous hysteresis in \smb, we hypothesize that the domain-wall conductance 
is equal to the anomalous quantum Hall conductance, 
and is quantized as 
$(n+1/2)e^{2}/h$, as expected in a massive Dirac spectrum induced by FM ordered moments pointing out of the surface plane \cite{qi08,hasan10}. The massive Dirac picture can explain the observed anisotropy of hysteresis in MR and WAL. For $H_{\perp}$ field orientation, surface Dirac electrons more polarized perpendicular to the conduction plane become more massive than for $H_{\perp}$ field orientation, leading to not only prominent hysteresis in MR by formation of the domain wall conduction, but also stronger suppression of the WAL by destruction of time reversal scattering processes in addition to suppression of spin-flip scattering \cite{lu11a}.

An observed surface state magnetoconductance approaching a value of $e^{2}/h$ reveals a key signature of quantized one-dimensional domain wall transport in \smb\ stabilized by Dirac electron-mediated ferromagnetism, and presents unequivocal evidence for the existence of topologically non-trivial surface states. The chiral modes on the surface states should be non-dissipative, but in real systems at finite temperatures, abundant inelastic scattering due to electron correlations \cite{konig07} or puddles due to spacial variation of the electronic structure \cite{marti08} easily suppress the ballistic quantum transport, leading to disappearance of quantized conductance. Nevertheless, we observe the quantized increase in the magnetoconductance, which indicates the formation of a grid-like FM domain structure. Provided the domain size is sufficiently smaller than the sample size of $\sim$ mm-scale, the domain walls effectively form an infinite resistor (conductor) network with components of $R=h/e^2$ ($G=e^2/h$) as shown in Fig.~3e. In such a network, the resistance between any two non-adjacent nodes is of the order $\sim R$, yielding the measured quantized conductance. 

The resistance component $R$ of the chiral modes is determined by the chemical potential of the Dirac electrons because multiple quantized channels with $(n+1/2)e^{2}/h$ could contribute to the conduction in {\smb}. A half-quantized conductance of $\frac{1}{2} e^{2}/h$ would be expected if the surface state Fermi energy is in the gap and in the lowest Landau level of $n=0$. Assuming both top and bottom surfaces contribute equally to the total conductance, this gives the observed value of $e^{2}/h = 2 \times \frac{1}{2}e^{2}/h$. Note that three Dirac bands are calculated to reside at the $\Gamma$ and $X/Y$ points in \smb\ \cite{xu13,neupa13,jiang13,frant13}.
The quantized conductance of $e^{2}/h$ suggest that the chemical potential sits in the gap only at the $\Gamma$ points, while it is above the gaps at the $X/Y$ points as shown in Fig.~3g. In fact, we observe the saturation of the resistance in this sample, which indicates that there remains surface conduction channels at very low temperatures as shown in the inset of Fig.~1a. In this case, a quantized anomalous Hall effect in the Hall resistance, such as observed in gate-tuned Cr-doped Bi$_{2}$Se$_{3}$ \cite{chang13} and BiSbTeSe$_2$ \cite{xu14}, is likely masked by the conduction of Dirac electrons at the $X/Y$ points. Finally, we note that the abrupt, sharp transitions observed in MR upon downsweep (Fig.~1b inset) often exhibit a jump in conductance very close to $\frac{1}{2}e^2/h$; whether this indicates a true quantization, or a unique signature of a chiral domain wall reconfiguration, remains a provocative observation to be explained.
With a truly insulating bulk band structure, future gating experiments \cite{syers15} utilizing single-crystal surfaces of \smb\ should readily facilitate the observation of these and other quantized properties in this system.

\bibliographystyle{apsrev4-1}
\bibliography{smb6_resubmit}

\begin{thebibliography}{39}%
\makeatletter
\providecommand \@ifxundefined [1]{%
 \@ifx{#1\undefined}
}%
\providecommand \@ifnum [1]{%
 \ifnum #1\expandafter \@firstoftwo
 \else \expandafter \@secondoftwo
 \fi
}%
\providecommand \@ifx [1]{%
 \ifx #1\expandafter \@firstoftwo
 \else \expandafter \@secondoftwo
 \fi
}%
\providecommand \natexlab [1]{#1}%
\providecommand \enquote  [1]{``#1''}%
\providecommand \bibnamefont  [1]{#1}%
\providecommand \bibfnamefont [1]{#1}%
\providecommand \citenamefont [1]{#1}%
\providecommand \href@noop [0]{\@secondoftwo}%
\providecommand \href [0]{\begingroup \@sanitize@url \@href}%
\providecommand \@href[1]{\@@startlink{#1}\@@href}%
\providecommand \@@href[1]{\endgroup#1\@@endlink}%
\providecommand \@sanitize@url [0]{\catcode `\\12\catcode `\$12\catcode
  `\&12\catcode `\#12\catcode `\^12\catcode `\_12\catcode `\%12\relax}%
\providecommand \@@startlink[1]{}%
\providecommand \@@endlink[0]{}%
\providecommand \url  [0]{\begingroup\@sanitize@url \@url }%
\providecommand \@url [1]{\endgroup\@href {#1}{\urlprefix }}%
\providecommand \urlprefix  [0]{URL }%
\providecommand \Eprint [0]{\href }%
\providecommand \doibase [0]{http://dx.doi.org/}%
\providecommand \selectlanguage [0]{\@gobble}%
\providecommand \bibinfo  [0]{\@secondoftwo}%
\providecommand \bibfield  [0]{\@secondoftwo}%
\providecommand \translation [1]{[#1]}%
\providecommand \BibitemOpen [0]{}%
\providecommand \bibitemStop [0]{}%
\providecommand \bibitemNoStop [0]{.\EOS\space}%
\providecommand \EOS [0]{\spacefactor3000\relax}%
\providecommand \BibitemShut  [1]{\csname bibitem#1\endcsname}%
\let\auto@bib@innerbib\@empty
\bibitem [{\citenamefont {Hasan}\ and\ \citenamefont {Kane}(2010)}]{hasan10}%
  \BibitemOpen
  \bibfield  {author} {\bibinfo {author} {\bibfnamefont {M.~Z.}\ \bibnamefont
  {Hasan}}\ and\ \bibinfo {author} {\bibfnamefont {C.~L.}\ \bibnamefont
  {Kane}},\ }\href {\doibase 10.1103/RevModPhys.82.3045} {\bibfield  {journal}
  {\bibinfo  {journal} {Rev. Mod. Phys.}\ }\textbf {\bibinfo {volume} {82}},\
  \bibinfo {pages} {3045} (\bibinfo {year} {2010})}\BibitemShut {NoStop}%
\bibitem [{\citenamefont {Wilczek}(2009)}]{wilcz09}%
  \BibitemOpen
  \bibfield  {author} {\bibinfo {author} {\bibfnamefont {F.}~\bibnamefont
  {Wilczek}},\ }\href@noop {} {\bibfield  {journal} {\bibinfo  {journal} {Nat.
  Phys.}\ }\textbf {\bibinfo {volume} {5}},\ \bibinfo {pages} {614} (\bibinfo
  {year} {2009})}\BibitemShut {NoStop}%
\bibitem [{\citenamefont {Wilczek}(1987)}]{wilcz87}%
  \BibitemOpen
  \bibfield  {author} {\bibinfo {author} {\bibfnamefont {F.}~\bibnamefont
  {Wilczek}},\ }\href {\doibase 10.1103/PhysRevLett.58.1799} {\bibfield
  {journal} {\bibinfo  {journal} {Phys. Rev. Lett.}\ }\textbf {\bibinfo
  {volume} {58}},\ \bibinfo {pages} {1799} (\bibinfo {year}
  {1987})}\BibitemShut {NoStop}%
\bibitem [{\citenamefont {Dzero}\ \emph {et~al.}(2010)\citenamefont {Dzero},
  \citenamefont {Sun}, \citenamefont {Galitski},\ and\ \citenamefont
  {Coleman}}]{dzero10}%
  \BibitemOpen
  \bibfield  {author} {\bibinfo {author} {\bibfnamefont {M.}~\bibnamefont
  {Dzero}}, \bibinfo {author} {\bibfnamefont {K.}~\bibnamefont {Sun}}, \bibinfo
  {author} {\bibfnamefont {V.}~\bibnamefont {Galitski}}, \ and\ \bibinfo
  {author} {\bibfnamefont {P.}~\bibnamefont {Coleman}},\ }\href {\doibase
  10.1103/PhysRevLett.104.106408} {\bibfield  {journal} {\bibinfo  {journal}
  {Phys. Rev. Lett.}\ }\textbf {\bibinfo {volume} {104}},\ \bibinfo {pages}
  {106408} (\bibinfo {year} {2010})}\BibitemShut {NoStop}%
\bibitem [{\citenamefont {Takimoto}(2011)}]{takim11}%
  \BibitemOpen
  \bibfield  {author} {\bibinfo {author} {\bibfnamefont {T.}~\bibnamefont
  {Takimoto}},\ }\href {\doibase 10.1143/JPSJ.80.123710} {\bibfield  {journal}
  {\bibinfo  {journal} {J. Phys. Soc. Jpn.}\ }\textbf {\bibinfo {volume}
  {80}},\ \bibinfo {pages} {123710} (\bibinfo {year} {2011})}\BibitemShut
  {NoStop}%
\bibitem [{\citenamefont {Menth}\ \emph {et~al.}(1969)\citenamefont {Menth},
  \citenamefont {Buehler},\ and\ \citenamefont {Geballe}}]{menth69}%
  \BibitemOpen
  \bibfield  {author} {\bibinfo {author} {\bibfnamefont {A.}~\bibnamefont
  {Menth}}, \bibinfo {author} {\bibfnamefont {E.}~\bibnamefont {Buehler}}, \
  and\ \bibinfo {author} {\bibfnamefont {T.~H.}\ \bibnamefont {Geballe}},\
  }\href {\doibase 10.1103/PhysRevLett.22.295} {\bibfield  {journal} {\bibinfo
  {journal} {Phys. Rev. Lett.}\ }\textbf {\bibinfo {volume} {22}},\ \bibinfo
  {pages} {295} (\bibinfo {year} {1969})}\BibitemShut {NoStop}%
\bibitem [{\citenamefont {Ando}(2013)}]{ando13}%
  \BibitemOpen
  \bibfield  {author} {\bibinfo {author} {\bibfnamefont {Y.}~\bibnamefont
  {Ando}},\ }\href {\doibase 10.1143/JPSJ.82.102001} {\bibfield  {journal}
  {\bibinfo  {journal} {J. Phys. Soc. Jpn.}\ }\textbf {\bibinfo {volume}
  {82}},\ \bibinfo {pages} {102001} (\bibinfo {year} {2013})}\BibitemShut
  {NoStop}%
\bibitem [{\citenamefont {Wolgast}\ \emph {et~al.}(2013)\citenamefont
  {Wolgast}, \citenamefont {Kurdak}, \citenamefont {Sun}, \citenamefont
  {Allen}, \citenamefont {Kim},\ and\ \citenamefont {Fisk}}]{wolga13}%
  \BibitemOpen
  \bibfield  {author} {\bibinfo {author} {\bibfnamefont {S.}~\bibnamefont
  {Wolgast}}, \bibinfo {author} {\bibfnamefont {i.~m. c. b. u. i. e. i.~f.}\
  \bibnamefont {Kurdak}}, \bibinfo {author} {\bibfnamefont {K.}~\bibnamefont
  {Sun}}, \bibinfo {author} {\bibfnamefont {J.~W.}\ \bibnamefont {Allen}},
  \bibinfo {author} {\bibfnamefont {D.-J.}\ \bibnamefont {Kim}}, \ and\
  \bibinfo {author} {\bibfnamefont {Z.}~\bibnamefont {Fisk}},\ }\href {\doibase
  10.1103/PhysRevB.88.180405} {\bibfield  {journal} {\bibinfo  {journal} {Phys.
  Rev. B}\ }\textbf {\bibinfo {volume} {88}},\ \bibinfo {pages} {180405}
  (\bibinfo {year} {2013})}\BibitemShut {NoStop}%
\bibitem [{\citenamefont {Kim}\ \emph {et~al.}(2013)\citenamefont {Kim},
  \citenamefont {Thomas}, \citenamefont {Grant}, \citenamefont {Botimer},
  \citenamefont {Fisk},\ and\ \citenamefont {Xia}}]{kim13a}%
  \BibitemOpen
  \bibfield  {author} {\bibinfo {author} {\bibfnamefont {D.~J.}\ \bibnamefont
  {Kim}}, \bibinfo {author} {\bibfnamefont {S.}~\bibnamefont {Thomas}},
  \bibinfo {author} {\bibfnamefont {T.}~\bibnamefont {Grant}}, \bibinfo
  {author} {\bibfnamefont {J.}~\bibnamefont {Botimer}}, \bibinfo {author}
  {\bibfnamefont {Z.}~\bibnamefont {Fisk}}, \ and\ \bibinfo {author}
  {\bibfnamefont {J.}~\bibnamefont {Xia}},\ }\href
  {http://dx.doi.org/10.1038/srep03150} {\bibfield  {journal} {\bibinfo
  {journal} {Sci. Rep.}\ }\textbf {\bibinfo {volume} {3}},\ \bibinfo {pages}
  {3150} (\bibinfo {year} {2013})}\BibitemShut {NoStop}%
\bibitem [{\citenamefont {Zhang}\ \emph {et~al.}(2013)\citenamefont {Zhang},
  \citenamefont {Butch}, \citenamefont {Syers}, \citenamefont {Ziemak},
  \citenamefont {Greene},\ and\ \citenamefont {Paglione}}]{zhang13}%
  \BibitemOpen
  \bibfield  {author} {\bibinfo {author} {\bibfnamefont {X.}~\bibnamefont
  {Zhang}}, \bibinfo {author} {\bibfnamefont {N.~P.}\ \bibnamefont {Butch}},
  \bibinfo {author} {\bibfnamefont {P.}~\bibnamefont {Syers}}, \bibinfo
  {author} {\bibfnamefont {S.}~\bibnamefont {Ziemak}}, \bibinfo {author}
  {\bibfnamefont {R.~L.}\ \bibnamefont {Greene}}, \ and\ \bibinfo {author}
  {\bibfnamefont {J.}~\bibnamefont {Paglione}},\ }\href {\doibase
  10.1103/PhysRevX.3.011011} {\bibfield  {journal} {\bibinfo  {journal} {Phys.
  Rev. X}\ }\textbf {\bibinfo {volume} {3}},\ \bibinfo {pages} {011011}
  (\bibinfo {year} {2013})}\BibitemShut {NoStop}%
\bibitem [{\citenamefont {Xu}\ \emph {et~al.}(2013)\citenamefont {Xu},
  \citenamefont {Shi}, \citenamefont {Biswas}, \citenamefont {Matt},
  \citenamefont {Dhaka}, \citenamefont {Huang}, \citenamefont {Plumb},
  \citenamefont {Radovi\ifmmode~\acute{c}\else \'{c}\fi{}}, \citenamefont
  {Dil}, \citenamefont {Pomjakushina}, \citenamefont {Conder}, \citenamefont
  {Amato}, \citenamefont {Salman}, \citenamefont {Paul}, \citenamefont {Mesot},
  \citenamefont {Ding},\ and\ \citenamefont {Shi}}]{xu13}%
  \BibitemOpen
  \bibfield  {author} {\bibinfo {author} {\bibfnamefont {N.}~\bibnamefont
  {Xu}}, \bibinfo {author} {\bibfnamefont {X.}~\bibnamefont {Shi}}, \bibinfo
  {author} {\bibfnamefont {P.~K.}\ \bibnamefont {Biswas}}, \bibinfo {author}
  {\bibfnamefont {C.~E.}\ \bibnamefont {Matt}}, \bibinfo {author}
  {\bibfnamefont {R.~S.}\ \bibnamefont {Dhaka}}, \bibinfo {author}
  {\bibfnamefont {Y.}~\bibnamefont {Huang}}, \bibinfo {author} {\bibfnamefont
  {N.~C.}\ \bibnamefont {Plumb}}, \bibinfo {author} {\bibfnamefont
  {M.}~\bibnamefont {Radovi\ifmmode~\acute{c}\else \'{c}\fi{}}}, \bibinfo
  {author} {\bibfnamefont {J.~H.}\ \bibnamefont {Dil}}, \bibinfo {author}
  {\bibfnamefont {E.}~\bibnamefont {Pomjakushina}}, \bibinfo {author}
  {\bibfnamefont {K.}~\bibnamefont {Conder}}, \bibinfo {author} {\bibfnamefont
  {A.}~\bibnamefont {Amato}}, \bibinfo {author} {\bibfnamefont
  {Z.}~\bibnamefont {Salman}}, \bibinfo {author} {\bibfnamefont {D.~M.}\
  \bibnamefont {Paul}}, \bibinfo {author} {\bibfnamefont {J.}~\bibnamefont
  {Mesot}}, \bibinfo {author} {\bibfnamefont {H.}~\bibnamefont {Ding}}, \ and\
  \bibinfo {author} {\bibfnamefont {M.}~\bibnamefont {Shi}},\ }\href {\doibase
  10.1103/PhysRevB.88.121102} {\bibfield  {journal} {\bibinfo  {journal} {Phys.
  Rev. B}\ }\textbf {\bibinfo {volume} {88}},\ \bibinfo {pages} {121102}
  (\bibinfo {year} {2013})}\BibitemShut {NoStop}%
\bibitem [{\citenamefont {Neupane}\ \emph {et~al.}(2013)\citenamefont
  {Neupane}, \citenamefont {Alidoust}, \citenamefont {Xu}, \citenamefont
  {Kondo}, \citenamefont {Ishida}, \citenamefont {Kim}, \citenamefont {Liu},
  \citenamefont {Belopolski}, \citenamefont {Jo}, \citenamefont {Chang},
  \citenamefont {Jeng}, \citenamefont {Durakiewicz}, \citenamefont {Balicas},
  \citenamefont {Lin}, \citenamefont {Bansil}, \citenamefont {Shin},
  \citenamefont {Fisk},\ and\ \citenamefont {Hasan}}]{neupa13}%
  \BibitemOpen
  \bibfield  {author} {\bibinfo {author} {\bibfnamefont {M.}~\bibnamefont
  {Neupane}}, \bibinfo {author} {\bibfnamefont {N.}~\bibnamefont {Alidoust}},
  \bibinfo {author} {\bibfnamefont {S.-Y.}\ \bibnamefont {Xu}}, \bibinfo
  {author} {\bibfnamefont {T.}~\bibnamefont {Kondo}}, \bibinfo {author}
  {\bibfnamefont {Y.}~\bibnamefont {Ishida}}, \bibinfo {author} {\bibfnamefont
  {D.~J.}\ \bibnamefont {Kim}}, \bibinfo {author} {\bibfnamefont
  {C.}~\bibnamefont {Liu}}, \bibinfo {author} {\bibfnamefont {I.}~\bibnamefont
  {Belopolski}}, \bibinfo {author} {\bibfnamefont {Y.~J.}\ \bibnamefont {Jo}},
  \bibinfo {author} {\bibfnamefont {T.-R.}\ \bibnamefont {Chang}}, \bibinfo
  {author} {\bibfnamefont {H.-T.}\ \bibnamefont {Jeng}}, \bibinfo {author}
  {\bibfnamefont {T.}~\bibnamefont {Durakiewicz}}, \bibinfo {author}
  {\bibfnamefont {L.}~\bibnamefont {Balicas}}, \bibinfo {author} {\bibfnamefont
  {H.}~\bibnamefont {Lin}}, \bibinfo {author} {\bibfnamefont {A.}~\bibnamefont
  {Bansil}}, \bibinfo {author} {\bibfnamefont {S.}~\bibnamefont {Shin}},
  \bibinfo {author} {\bibfnamefont {Z.}~\bibnamefont {Fisk}}, \ and\ \bibinfo
  {author} {\bibfnamefont {M.~Z.}\ \bibnamefont {Hasan}},\ }\href
  {http://dx.doi.org/10.1038/ncomms3991} {\bibfield  {journal} {\bibinfo
  {journal} {Nat. Commun.}\ }\textbf {\bibinfo {volume} {4}},\ \bibinfo {pages}
  {2991} (\bibinfo {year} {2013})}\BibitemShut {NoStop}%
\bibitem [{\citenamefont {Jiang}\ \emph {et~al.}(2013)\citenamefont {Jiang},
  \citenamefont {Li}, \citenamefont {Zhang}, \citenamefont {Sun}, \citenamefont
  {Chen}, \citenamefont {Ye}, \citenamefont {Xu}, \citenamefont {Ge},
  \citenamefont {Tan}, \citenamefont {Niu}, \citenamefont {Xia}, \citenamefont
  {Xie}, \citenamefont {Li}, \citenamefont {Chen}, \citenamefont {Wen},\ and\
  \citenamefont {Feng}}]{jiang13}%
  \BibitemOpen
  \bibfield  {author} {\bibinfo {author} {\bibfnamefont {J.}~\bibnamefont
  {Jiang}}, \bibinfo {author} {\bibfnamefont {S.}~\bibnamefont {Li}}, \bibinfo
  {author} {\bibfnamefont {T.}~\bibnamefont {Zhang}}, \bibinfo {author}
  {\bibfnamefont {Z.}~\bibnamefont {Sun}}, \bibinfo {author} {\bibfnamefont
  {F.}~\bibnamefont {Chen}}, \bibinfo {author} {\bibfnamefont {Z.~R.}\
  \bibnamefont {Ye}}, \bibinfo {author} {\bibfnamefont {M.}~\bibnamefont {Xu}},
  \bibinfo {author} {\bibfnamefont {Q.~Q.}\ \bibnamefont {Ge}}, \bibinfo
  {author} {\bibfnamefont {S.~Y.}\ \bibnamefont {Tan}}, \bibinfo {author}
  {\bibfnamefont {X.~H.}\ \bibnamefont {Niu}}, \bibinfo {author} {\bibfnamefont
  {M.}~\bibnamefont {Xia}}, \bibinfo {author} {\bibfnamefont {B.~P.}\
  \bibnamefont {Xie}}, \bibinfo {author} {\bibfnamefont {Y.~F.}\ \bibnamefont
  {Li}}, \bibinfo {author} {\bibfnamefont {X.~H.}\ \bibnamefont {Chen}},
  \bibinfo {author} {\bibfnamefont {H.~H.}\ \bibnamefont {Wen}}, \ and\
  \bibinfo {author} {\bibfnamefont {D.~L.}\ \bibnamefont {Feng}},\ }\href
  {http://dx.doi.org/10.1038/ncomms4010} {\bibfield  {journal} {\bibinfo
  {journal} {Nat Commun}\ }\textbf {\bibinfo {volume} {4}},\ \bibinfo {pages}
  {3010} (\bibinfo {year} {2013})}\BibitemShut {NoStop}%
\bibitem [{\citenamefont {Frantzeskakis}\ \emph {et~al.}(2013)\citenamefont
  {Frantzeskakis}, \citenamefont {de~Jong}, \citenamefont {Zwartsenberg},
  \citenamefont {Huang}, \citenamefont {Pan}, \citenamefont {Zhang},
  \citenamefont {Zhang}, \citenamefont {Zhang}, \citenamefont {Bao},
  \citenamefont {Tegus}, \citenamefont {Varykhalov}, \citenamefont
  {de~Visser},\ and\ \citenamefont {Golden}}]{frant13}%
  \BibitemOpen
  \bibfield  {author} {\bibinfo {author} {\bibfnamefont {E.}~\bibnamefont
  {Frantzeskakis}}, \bibinfo {author} {\bibfnamefont {N.}~\bibnamefont
  {de~Jong}}, \bibinfo {author} {\bibfnamefont {B.}~\bibnamefont
  {Zwartsenberg}}, \bibinfo {author} {\bibfnamefont {Y.~K.}\ \bibnamefont
  {Huang}}, \bibinfo {author} {\bibfnamefont {Y.}~\bibnamefont {Pan}}, \bibinfo
  {author} {\bibfnamefont {X.}~\bibnamefont {Zhang}}, \bibinfo {author}
  {\bibfnamefont {J.~X.}\ \bibnamefont {Zhang}}, \bibinfo {author}
  {\bibfnamefont {F.~X.}\ \bibnamefont {Zhang}}, \bibinfo {author}
  {\bibfnamefont {L.~H.}\ \bibnamefont {Bao}}, \bibinfo {author} {\bibfnamefont
  {O.}~\bibnamefont {Tegus}}, \bibinfo {author} {\bibfnamefont
  {A.}~\bibnamefont {Varykhalov}}, \bibinfo {author} {\bibfnamefont
  {A.}~\bibnamefont {de~Visser}}, \ and\ \bibinfo {author} {\bibfnamefont
  {M.~S.}\ \bibnamefont {Golden}},\ }\href {\doibase 10.1103/PhysRevX.3.041024}
  {\bibfield  {journal} {\bibinfo  {journal} {Phys. Rev. X}\ }\textbf {\bibinfo
  {volume} {3}},\ \bibinfo {pages} {041024} (\bibinfo {year}
  {2013})}\BibitemShut {NoStop}%
\bibitem [{\citenamefont {Zhu}\ \emph {et~al.}(2013)\citenamefont {Zhu},
  \citenamefont {Nicolaou}, \citenamefont {Levy}, \citenamefont {Butch},
  \citenamefont {Syers}, \citenamefont {Wang}, \citenamefont {Paglione},
  \citenamefont {Sawatzky}, \citenamefont {Elfimov},\ and\ \citenamefont
  {Damascelli}}]{zhu13}%
  \BibitemOpen
  \bibfield  {author} {\bibinfo {author} {\bibfnamefont {Z.-H.}\ \bibnamefont
  {Zhu}}, \bibinfo {author} {\bibfnamefont {A.}~\bibnamefont {Nicolaou}},
  \bibinfo {author} {\bibfnamefont {G.}~\bibnamefont {Levy}}, \bibinfo {author}
  {\bibfnamefont {N.~P.}\ \bibnamefont {Butch}}, \bibinfo {author}
  {\bibfnamefont {P.}~\bibnamefont {Syers}}, \bibinfo {author} {\bibfnamefont
  {X.~F.}\ \bibnamefont {Wang}}, \bibinfo {author} {\bibfnamefont
  {J.}~\bibnamefont {Paglione}}, \bibinfo {author} {\bibfnamefont {G.~A.}\
  \bibnamefont {Sawatzky}}, \bibinfo {author} {\bibfnamefont {I.~S.}\
  \bibnamefont {Elfimov}}, \ and\ \bibinfo {author} {\bibfnamefont
  {A.}~\bibnamefont {Damascelli}},\ }\href {\doibase
  10.1103/PhysRevLett.111.216402} {\bibfield  {journal} {\bibinfo  {journal}
  {Phys. Rev. Lett.}\ }\textbf {\bibinfo {volume} {111}},\ \bibinfo {pages}
  {216402} (\bibinfo {year} {2013})}\BibitemShut {NoStop}%
\bibitem [{\citenamefont {Kim}\ \emph {et~al.}(2014)\citenamefont {Kim},
  \citenamefont {Xia},\ and\ \citenamefont {Fisk}}]{kim14}%
  \BibitemOpen
  \bibfield  {author} {\bibinfo {author} {\bibfnamefont {D.~J.}\ \bibnamefont
  {Kim}}, \bibinfo {author} {\bibfnamefont {J.}~\bibnamefont {Xia}}, \ and\
  \bibinfo {author} {\bibfnamefont {Z.}~\bibnamefont {Fisk}},\ }\href
  {http://dx.doi.org/10.1038/nmat3913} {\bibfield  {journal} {\bibinfo
  {journal} {Nat Mater}\ }\textbf {\bibinfo {volume} {13}},\ \bibinfo {pages}
  {466} (\bibinfo {year} {2014})}\BibitemShut {NoStop}%
\bibitem [{\citenamefont {Thomas}\ \emph {et~al.}(2013)\citenamefont {Thomas},
  \citenamefont {Kim}, \citenamefont {Chung}, \citenamefont {Grant},
  \citenamefont {Fisk},\ and\ \citenamefont {Xia}}]{thoma13}%
  \BibitemOpen
  \bibfield  {author} {\bibinfo {author} {\bibfnamefont {S.}~\bibnamefont
  {Thomas}}, \bibinfo {author} {\bibfnamefont {D.}~\bibnamefont {Kim}},
  \bibinfo {author} {\bibfnamefont {S.~B.}\ \bibnamefont {Chung}}, \bibinfo
  {author} {\bibfnamefont {T.}~\bibnamefont {Grant}}, \bibinfo {author}
  {\bibfnamefont {Z.}~\bibnamefont {Fisk}}, \ and\ \bibinfo {author}
  {\bibfnamefont {J.}~\bibnamefont {Xia}},\ }\href@noop {} {\bibfield
  {journal} {\bibinfo  {journal} {arXiv:1307.4133}\ } (\bibinfo {year}
  {2013})}\BibitemShut {NoStop}%
\bibitem [{\citenamefont {Syers}\ \emph {et~al.}(2015)\citenamefont {Syers},
  \citenamefont {Kim}, \citenamefont {Fuhrer},\ and\ \citenamefont
  {Paglione}}]{syers15}%
  \BibitemOpen
  \bibfield  {author} {\bibinfo {author} {\bibfnamefont {P.}~\bibnamefont
  {Syers}}, \bibinfo {author} {\bibfnamefont {D.}~\bibnamefont {Kim}}, \bibinfo
  {author} {\bibfnamefont {M.~S.}\ \bibnamefont {Fuhrer}}, \ and\ \bibinfo
  {author} {\bibfnamefont {J.}~\bibnamefont {Paglione}},\ }\href {\doibase
  10.1103/PhysRevLett.114.096601} {\bibfield  {journal} {\bibinfo  {journal}
  {Phys. Rev. Lett.}\ }\textbf {\bibinfo {volume} {114}},\ \bibinfo {pages}
  {096601} (\bibinfo {year} {2015})}\BibitemShut {NoStop}%
\bibitem [{\citenamefont {Cooley}\ \emph
  {et~al.}(1995{\natexlab{a}})\citenamefont {Cooley}, \citenamefont {Aronson},
  \citenamefont {Lacerda}, \citenamefont {Fisk}, \citenamefont {Canfield},\
  and\ \citenamefont {Guertin}}]{coole95a}%
  \BibitemOpen
  \bibfield  {author} {\bibinfo {author} {\bibfnamefont {J.~C.}\ \bibnamefont
  {Cooley}}, \bibinfo {author} {\bibfnamefont {M.~C.}\ \bibnamefont {Aronson}},
  \bibinfo {author} {\bibfnamefont {A.}~\bibnamefont {Lacerda}}, \bibinfo
  {author} {\bibfnamefont {Z.}~\bibnamefont {Fisk}}, \bibinfo {author}
  {\bibfnamefont {P.~C.}\ \bibnamefont {Canfield}}, \ and\ \bibinfo {author}
  {\bibfnamefont {R.~P.}\ \bibnamefont {Guertin}},\ }\href {\doibase
  10.1103/PhysRevB.52.7322} {\bibfield  {journal} {\bibinfo  {journal} {Phys.
  Rev. B}\ }\textbf {\bibinfo {volume} {52}},\ \bibinfo {pages} {7322}
  (\bibinfo {year} {1995}{\natexlab{a}})}\BibitemShut {NoStop}%
\bibitem [{\citenamefont {Fu}\ and\ \citenamefont {Kane}(2007)}]{fu07a}%
  \BibitemOpen
  \bibfield  {author} {\bibinfo {author} {\bibfnamefont {L.}~\bibnamefont
  {Fu}}\ and\ \bibinfo {author} {\bibfnamefont {C.~L.}\ \bibnamefont {Kane}},\
  }\href {\doibase 10.1103/PhysRevB.76.045302} {\bibfield  {journal} {\bibinfo
  {journal} {Phys. Rev. B}\ }\textbf {\bibinfo {volume} {76}},\ \bibinfo
  {pages} {045302} (\bibinfo {year} {2007})}\BibitemShut {NoStop}%
\bibitem [{\citenamefont {Hikami}\ \emph {et~al.}(1980)\citenamefont {Hikami},
  \citenamefont {Larkin},\ and\ \citenamefont {Nagaoka}}]{hikam80}%
  \BibitemOpen
  \bibfield  {author} {\bibinfo {author} {\bibfnamefont {S.}~\bibnamefont
  {Hikami}}, \bibinfo {author} {\bibfnamefont {A.~I.}\ \bibnamefont {Larkin}},
  \ and\ \bibinfo {author} {\bibfnamefont {Y.}~\bibnamefont {Nagaoka}},\ }\href
  {\doibase 10.1143/PTP.63.707} {\bibfield  {journal} {\bibinfo  {journal}
  {Prog. Theor. Phys.}\ }\textbf {\bibinfo {volume} {63}},\ \bibinfo {pages}
  {707} (\bibinfo {year} {1980})}\BibitemShut {NoStop}%
\bibitem [{\citenamefont {Zhang}\ \emph {et~al.}(2010)\citenamefont {Zhang},
  \citenamefont {He}, \citenamefont {Chang}, \citenamefont {Song},
  \citenamefont {Wang}, \citenamefont {Chen}, \citenamefont {Jia},
  \citenamefont {Fang}, \citenamefont {Dai}, \citenamefont {Shan},
  \citenamefont {Shen}, \citenamefont {Niu}, \citenamefont {Qi}, \citenamefont
  {Zhang}, \citenamefont {Ma},\ and\ \citenamefont {Xue}}]{zhang10a}%
  \BibitemOpen
  \bibfield  {author} {\bibinfo {author} {\bibfnamefont {Y.}~\bibnamefont
  {Zhang}}, \bibinfo {author} {\bibfnamefont {K.}~\bibnamefont {He}}, \bibinfo
  {author} {\bibfnamefont {C.-Z.}\ \bibnamefont {Chang}}, \bibinfo {author}
  {\bibfnamefont {C.-L.}\ \bibnamefont {Song}}, \bibinfo {author}
  {\bibfnamefont {L.-L.}\ \bibnamefont {Wang}}, \bibinfo {author}
  {\bibfnamefont {X.}~\bibnamefont {Chen}}, \bibinfo {author} {\bibfnamefont
  {J.-F.}\ \bibnamefont {Jia}}, \bibinfo {author} {\bibfnamefont
  {Z.}~\bibnamefont {Fang}}, \bibinfo {author} {\bibfnamefont {X.}~\bibnamefont
  {Dai}}, \bibinfo {author} {\bibfnamefont {W.-Y.}\ \bibnamefont {Shan}},
  \bibinfo {author} {\bibfnamefont {S.-Q.}\ \bibnamefont {Shen}}, \bibinfo
  {author} {\bibfnamefont {Q.}~\bibnamefont {Niu}}, \bibinfo {author}
  {\bibfnamefont {X.-L.}\ \bibnamefont {Qi}}, \bibinfo {author} {\bibfnamefont
  {S.-C.}\ \bibnamefont {Zhang}}, \bibinfo {author} {\bibfnamefont {X.-C.}\
  \bibnamefont {Ma}}, \ and\ \bibinfo {author} {\bibfnamefont {Q.-K.}\
  \bibnamefont {Xue}},\ }\href {http://dx.doi.org/10.1038/nphys1689} {\bibfield
   {journal} {\bibinfo  {journal} {Nat Phys}\ }\textbf {\bibinfo {volume}
  {6}},\ \bibinfo {pages} {584} (\bibinfo {year} {2010})}\BibitemShut {NoStop}%
\bibitem [{\citenamefont {Cooley}\ \emph
  {et~al.}(1995{\natexlab{b}})\citenamefont {Cooley}, \citenamefont {Aronson},
  \citenamefont {Fisk},\ and\ \citenamefont {Canfield}}]{coole95}%
  \BibitemOpen
  \bibfield  {author} {\bibinfo {author} {\bibfnamefont {J.~C.}\ \bibnamefont
  {Cooley}}, \bibinfo {author} {\bibfnamefont {M.~C.}\ \bibnamefont {Aronson}},
  \bibinfo {author} {\bibfnamefont {Z.}~\bibnamefont {Fisk}}, \ and\ \bibinfo
  {author} {\bibfnamefont {P.~C.}\ \bibnamefont {Canfield}},\ }\href {\doibase
  10.1103/PhysRevLett.74.1629} {\bibfield  {journal} {\bibinfo  {journal}
  {Phys. Rev. Lett.}\ }\textbf {\bibinfo {volume} {74}},\ \bibinfo {pages}
  {1629} (\bibinfo {year} {1995}{\natexlab{b}})}\BibitemShut {NoStop}%
\bibitem [{\citenamefont {Tkachov}\ and\ \citenamefont
  {Hankiewicz}(2011)}]{tkach11}%
  \BibitemOpen
  \bibfield  {author} {\bibinfo {author} {\bibfnamefont {G.}~\bibnamefont
  {Tkachov}}\ and\ \bibinfo {author} {\bibfnamefont {E.~M.}\ \bibnamefont
  {Hankiewicz}},\ }\href {\doibase 10.1103/PhysRevB.84.035444} {\bibfield
  {journal} {\bibinfo  {journal} {Phys. Rev. B}\ }\textbf {\bibinfo {volume}
  {84}},\ \bibinfo {pages} {035444} (\bibinfo {year} {2011})}\BibitemShut
  {NoStop}%
\bibitem [{\citenamefont {Hamidian}\ \emph {et~al.}(2011)\citenamefont
  {Hamidian}, \citenamefont {Schmidt}, \citenamefont {Firmo}, \citenamefont
  {Allan}, \citenamefont {Bradley}, \citenamefont {Garrett}, \citenamefont
  {Williams}, \citenamefont {Luke}, \citenamefont {Dubi}, \citenamefont
  {Balatsky},\ and\ \citenamefont {Davis}}]{hamid11}%
  \BibitemOpen
  \bibfield  {author} {\bibinfo {author} {\bibfnamefont {M.~H.}\ \bibnamefont
  {Hamidian}}, \bibinfo {author} {\bibfnamefont {A.~R.}\ \bibnamefont
  {Schmidt}}, \bibinfo {author} {\bibfnamefont {I.~A.}\ \bibnamefont {Firmo}},
  \bibinfo {author} {\bibfnamefont {M.~P.}\ \bibnamefont {Allan}}, \bibinfo
  {author} {\bibfnamefont {P.}~\bibnamefont {Bradley}}, \bibinfo {author}
  {\bibfnamefont {J.~D.}\ \bibnamefont {Garrett}}, \bibinfo {author}
  {\bibfnamefont {T.~J.}\ \bibnamefont {Williams}}, \bibinfo {author}
  {\bibfnamefont {G.~M.}\ \bibnamefont {Luke}}, \bibinfo {author}
  {\bibfnamefont {Y.}~\bibnamefont {Dubi}}, \bibinfo {author} {\bibfnamefont
  {A.~V.}\ \bibnamefont {Balatsky}}, \ and\ \bibinfo {author} {\bibfnamefont
  {J.~C.}\ \bibnamefont {Davis}},\ }\href {\doibase 10.1073/pnas.1115027108}
  {\bibfield  {journal} {\bibinfo  {journal} {Proc. Natl. Acad. Sci.}\ }\textbf
  {\bibinfo {volume} {108}},\ \bibinfo {pages} {18233} (\bibinfo {year}
  {2011})},\ \Eprint
  {http://arxiv.org/abs/http://www.pnas.org/content/108/45/18233.full.pdf+html}
  {http://www.pnas.org/content/108/45/18233.full.pdf+html} \BibitemShut
  {NoStop}%
\bibitem [{\citenamefont {Phelan}\ \emph {et~al.}(2014)\citenamefont {Phelan},
  \citenamefont {Koohpayeh}, \citenamefont {Cottingham}, \citenamefont
  {Freeland}, \citenamefont {Leiner}, \citenamefont {Broholm},\ and\
  \citenamefont {McQueen}}]{phela14}%
  \BibitemOpen
  \bibfield  {author} {\bibinfo {author} {\bibfnamefont {W.~A.}\ \bibnamefont
  {Phelan}}, \bibinfo {author} {\bibfnamefont {S.~M.}\ \bibnamefont
  {Koohpayeh}}, \bibinfo {author} {\bibfnamefont {P.}~\bibnamefont
  {Cottingham}}, \bibinfo {author} {\bibfnamefont {J.~W.}\ \bibnamefont
  {Freeland}}, \bibinfo {author} {\bibfnamefont {J.~C.}\ \bibnamefont
  {Leiner}}, \bibinfo {author} {\bibfnamefont {C.~L.}\ \bibnamefont {Broholm}},
  \ and\ \bibinfo {author} {\bibfnamefont {T.~M.}\ \bibnamefont {McQueen}},\
  }\href {\doibase 10.1103/PhysRevX.4.031012} {\bibfield  {journal} {\bibinfo
  {journal} {Phys. Rev. X}\ }\textbf {\bibinfo {volume} {4}},\ \bibinfo {pages}
  {031012} (\bibinfo {year} {2014})}\BibitemShut {NoStop}%
\bibitem [{\citenamefont {Liu}\ \emph {et~al.}(2009)\citenamefont {Liu},
  \citenamefont {Liu}, \citenamefont {Xu}, \citenamefont {Qi},\ and\
  \citenamefont {Zhang}}]{liu09a}%
  \BibitemOpen
  \bibfield  {author} {\bibinfo {author} {\bibfnamefont {Q.}~\bibnamefont
  {Liu}}, \bibinfo {author} {\bibfnamefont {C.-X.}\ \bibnamefont {Liu}},
  \bibinfo {author} {\bibfnamefont {C.}~\bibnamefont {Xu}}, \bibinfo {author}
  {\bibfnamefont {X.-L.}\ \bibnamefont {Qi}}, \ and\ \bibinfo {author}
  {\bibfnamefont {S.-C.}\ \bibnamefont {Zhang}},\ }\href {\doibase
  10.1103/PhysRevLett.102.156603} {\bibfield  {journal} {\bibinfo  {journal}
  {Phys. Rev. Lett.}\ }\textbf {\bibinfo {volume} {102}},\ \bibinfo {pages}
  {156603} (\bibinfo {year} {2009})}\BibitemShut {NoStop}%
\bibitem [{\citenamefont {Nagaosa}\ \emph {et~al.}(2010)\citenamefont
  {Nagaosa}, \citenamefont {Sinova}, \citenamefont {Onoda}, \citenamefont
  {MacDonald},\ and\ \citenamefont {Ong}}]{nagao10}%
  \BibitemOpen
  \bibfield  {author} {\bibinfo {author} {\bibfnamefont {N.}~\bibnamefont
  {Nagaosa}}, \bibinfo {author} {\bibfnamefont {J.}~\bibnamefont {Sinova}},
  \bibinfo {author} {\bibfnamefont {S.}~\bibnamefont {Onoda}}, \bibinfo
  {author} {\bibfnamefont {A.~H.}\ \bibnamefont {MacDonald}}, \ and\ \bibinfo
  {author} {\bibfnamefont {N.~P.}\ \bibnamefont {Ong}},\ }\href {\doibase
  10.1103/RevModPhys.82.1539} {\bibfield  {journal} {\bibinfo  {journal} {Rev.
  Mod. Phys.}\ }\textbf {\bibinfo {volume} {82}},\ \bibinfo {pages} {1539}
  (\bibinfo {year} {2010})}\BibitemShut {NoStop}%
\bibitem [{\citenamefont {Allen}\ \emph {et~al.}(1979)\citenamefont {Allen},
  \citenamefont {Batlogg},\ and\ \citenamefont {Wachter}}]{allen79}%
  \BibitemOpen
  \bibfield  {author} {\bibinfo {author} {\bibfnamefont {J.~W.}\ \bibnamefont
  {Allen}}, \bibinfo {author} {\bibfnamefont {B.}~\bibnamefont {Batlogg}}, \
  and\ \bibinfo {author} {\bibfnamefont {P.}~\bibnamefont {Wachter}},\ }\href
  {\doibase 10.1103/PhysRevB.20.4807} {\bibfield  {journal} {\bibinfo
  {journal} {Phys. Rev. B}\ }\textbf {\bibinfo {volume} {20}},\ \bibinfo
  {pages} {4807} (\bibinfo {year} {1979})}\BibitemShut {NoStop}%
\bibitem [{\citenamefont {Giri}\ \emph {et~al.}(2011)\citenamefont {Giri},
  \citenamefont {Patra},\ and\ \citenamefont {Majumdar}}]{giri11}%
  \BibitemOpen
  \bibfield  {author} {\bibinfo {author} {\bibfnamefont {S.}~\bibnamefont
  {Giri}}, \bibinfo {author} {\bibfnamefont {M.}~\bibnamefont {Patra}}, \ and\
  \bibinfo {author} {\bibfnamefont {S.}~\bibnamefont {Majumdar}},\ }\href
  {http://stacks.iop.org/0953-8984/23/i=7/a=073201} {\bibfield  {journal}
  {\bibinfo  {journal} {J. Phys: Condens. Matter}\ }\textbf {\bibinfo {volume}
  {23}},\ \bibinfo {pages} {073201} (\bibinfo {year} {2011})}\BibitemShut
  {NoStop}%
\bibitem [{\citenamefont {McCord}\ and\ \citenamefont
  {Sch{\"a}fer}(2009)}]{mccor09}%
  \BibitemOpen
  \bibfield  {author} {\bibinfo {author} {\bibfnamefont {J.}~\bibnamefont
  {McCord}}\ and\ \bibinfo {author} {\bibfnamefont {R.}~\bibnamefont
  {Sch{\"a}fer}},\ }\href {http://stacks.iop.org/1367-2630/11/i=8/a=083016}
  {\bibfield  {journal} {\bibinfo  {journal} {New Journal of Physics}\ }\textbf
  {\bibinfo {volume} {11}},\ \bibinfo {pages} {083016} (\bibinfo {year}
  {2009})}\BibitemShut {NoStop}%
\bibitem [{\citenamefont {Esho}(1976)}]{esho76}%
  \BibitemOpen
  \bibfield  {author} {\bibinfo {author} {\bibfnamefont {S.}~\bibnamefont
  {Esho}},\ }\href {\doibase 10.7567/JJAPS.15S1.93} {\bibfield  {journal}
  {\bibinfo  {journal} {Jpn. J. Appl. Phys.}\ }\textbf {\bibinfo {volume}
  {15S1}},\ \bibinfo {pages} {93} (\bibinfo {year} {1976})}\BibitemShut
  {NoStop}%
\bibitem [{\citenamefont {Checkelsky}\ \emph {et~al.}(2012)\citenamefont
  {Checkelsky}, \citenamefont {Ye}, \citenamefont {Onose}, \citenamefont
  {Iwasa},\ and\ \citenamefont {Tokura}}]{check12}%
  \BibitemOpen
  \bibfield  {author} {\bibinfo {author} {\bibfnamefont {J.~G.}\ \bibnamefont
  {Checkelsky}}, \bibinfo {author} {\bibfnamefont {J.}~\bibnamefont {Ye}},
  \bibinfo {author} {\bibfnamefont {Y.}~\bibnamefont {Onose}}, \bibinfo
  {author} {\bibfnamefont {Y.}~\bibnamefont {Iwasa}}, \ and\ \bibinfo {author}
  {\bibfnamefont {Y.}~\bibnamefont {Tokura}},\ }\href@noop {} {\bibfield
  {journal} {\bibinfo  {journal} {Nat. Phys.}\ }\textbf {\bibinfo {volume}
  {8}},\ \bibinfo {pages} {729} (\bibinfo {year} {2012})}\BibitemShut {NoStop}%
\bibitem [{\citenamefont {Qi}\ \emph {et~al.}(2008)\citenamefont {Qi},
  \citenamefont {Hughes},\ and\ \citenamefont {Zhang}}]{qi08}%
  \BibitemOpen
  \bibfield  {author} {\bibinfo {author} {\bibfnamefont {X.-L.}\ \bibnamefont
  {Qi}}, \bibinfo {author} {\bibfnamefont {T.~L.}\ \bibnamefont {Hughes}}, \
  and\ \bibinfo {author} {\bibfnamefont {S.-C.}\ \bibnamefont {Zhang}},\ }\href
  {\doibase 10.1103/PhysRevB.78.195424} {\bibfield  {journal} {\bibinfo
  {journal} {Phys. Rev. B}\ }\textbf {\bibinfo {volume} {78}},\ \bibinfo {eid}
  {195424} (\bibinfo {year} {2008})}\BibitemShut {NoStop}%
\bibitem [{\citenamefont {Lu}\ and\ \citenamefont {Shen}(2011)}]{lu11a}%
  \BibitemOpen
  \bibfield  {author} {\bibinfo {author} {\bibfnamefont {H.-Z.}\ \bibnamefont
  {Lu}}\ and\ \bibinfo {author} {\bibfnamefont {S.-Q.}\ \bibnamefont {Shen}},\
  }\href {\doibase 10.1103/PhysRevB.84.125138} {\bibfield  {journal} {\bibinfo
  {journal} {Phys. Rev. B}\ }\textbf {\bibinfo {volume} {84}},\ \bibinfo
  {pages} {125138} (\bibinfo {year} {2011})}\BibitemShut {NoStop}%
\bibitem [{\citenamefont {Konig}\ \emph {et~al.}(2007)\citenamefont {Konig},
  \citenamefont {Wiedmann}, \citenamefont {Brune}, \citenamefont {Roth},
  \citenamefont {Buhmann}, \citenamefont {Molenkamp}, \citenamefont {Qi},\ and\
  \citenamefont {Zhang}}]{konig07}%
  \BibitemOpen
  \bibfield  {author} {\bibinfo {author} {\bibfnamefont {M.}~\bibnamefont
  {Konig}}, \bibinfo {author} {\bibfnamefont {S.}~\bibnamefont {Wiedmann}},
  \bibinfo {author} {\bibfnamefont {C.}~\bibnamefont {Brune}}, \bibinfo
  {author} {\bibfnamefont {A.}~\bibnamefont {Roth}}, \bibinfo {author}
  {\bibfnamefont {H.}~\bibnamefont {Buhmann}}, \bibinfo {author} {\bibfnamefont
  {L.~W.}\ \bibnamefont {Molenkamp}}, \bibinfo {author} {\bibfnamefont {X.-L.}\
  \bibnamefont {Qi}}, \ and\ \bibinfo {author} {\bibfnamefont {S.-C.}\
  \bibnamefont {Zhang}},\ }\href {\doibase 10.1126/science.1148047} {\bibfield
  {journal} {\bibinfo  {journal} {Science}\ }\textbf {\bibinfo {volume}
  {318}},\ \bibinfo {pages} {766} (\bibinfo {year} {2007})},\ \Eprint
  {http://arxiv.org/abs/http://www.sciencemag.org/cgi/reprint/318/5851/766.pdf}
  {http://www.sciencemag.org/cgi/reprint/318/5851/766.pdf} \BibitemShut
  {NoStop}%
\bibitem [{\citenamefont {Martin}\ \emph {et~al.}(2008)\citenamefont {Martin},
  \citenamefont {Akerman}, \citenamefont {Ulbricht}, \citenamefont {Lohmann},
  \citenamefont {Smet}, \citenamefont {von Klitzing},\ and\ \citenamefont
  {Yacoby}}]{marti08}%
  \BibitemOpen
  \bibfield  {author} {\bibinfo {author} {\bibfnamefont {J.}~\bibnamefont
  {Martin}}, \bibinfo {author} {\bibfnamefont {N.}~\bibnamefont {Akerman}},
  \bibinfo {author} {\bibfnamefont {G.}~\bibnamefont {Ulbricht}}, \bibinfo
  {author} {\bibfnamefont {T.}~\bibnamefont {Lohmann}}, \bibinfo {author}
  {\bibfnamefont {J.~H.}\ \bibnamefont {Smet}}, \bibinfo {author}
  {\bibfnamefont {K.}~\bibnamefont {von Klitzing}}, \ and\ \bibinfo {author}
  {\bibfnamefont {A.}~\bibnamefont {Yacoby}},\ }\href
  {http://dx.doi.org/10.1038/nphys781} {\bibfield  {journal} {\bibinfo
  {journal} {Nat Phys}\ }\textbf {\bibinfo {volume} {4}},\ \bibinfo {pages}
  {144} (\bibinfo {year} {2008})}\BibitemShut {NoStop}%
\bibitem [{\citenamefont {Chang}\ \emph {et~al.}(2013)\citenamefont {Chang},
  \citenamefont {Zhang}, \citenamefont {Feng}, \citenamefont {Shen},
  \citenamefont {Zhang}, \citenamefont {Guo}, \citenamefont {Li}, \citenamefont
  {Ou}, \citenamefont {Wei}, \citenamefont {Wang}, \citenamefont {Ji},
  \citenamefont {Feng}, \citenamefont {Ji}, \citenamefont {Chen}, \citenamefont
  {Jia}, \citenamefont {Dai}, \citenamefont {Fang}, \citenamefont {Zhang},
  \citenamefont {He}, \citenamefont {Wang}, \citenamefont {Lu}, \citenamefont
  {Ma},\ and\ \citenamefont {Xue}}]{chang13}%
  \BibitemOpen
  \bibfield  {author} {\bibinfo {author} {\bibfnamefont {C.-Z.}\ \bibnamefont
  {Chang}}, \bibinfo {author} {\bibfnamefont {J.}~\bibnamefont {Zhang}},
  \bibinfo {author} {\bibfnamefont {X.}~\bibnamefont {Feng}}, \bibinfo {author}
  {\bibfnamefont {J.}~\bibnamefont {Shen}}, \bibinfo {author} {\bibfnamefont
  {Z.}~\bibnamefont {Zhang}}, \bibinfo {author} {\bibfnamefont
  {M.}~\bibnamefont {Guo}}, \bibinfo {author} {\bibfnamefont {K.}~\bibnamefont
  {Li}}, \bibinfo {author} {\bibfnamefont {Y.}~\bibnamefont {Ou}}, \bibinfo
  {author} {\bibfnamefont {P.}~\bibnamefont {Wei}}, \bibinfo {author}
  {\bibfnamefont {L.-L.}\ \bibnamefont {Wang}}, \bibinfo {author}
  {\bibfnamefont {Z.-Q.}\ \bibnamefont {Ji}}, \bibinfo {author} {\bibfnamefont
  {Y.}~\bibnamefont {Feng}}, \bibinfo {author} {\bibfnamefont {S.}~\bibnamefont
  {Ji}}, \bibinfo {author} {\bibfnamefont {X.}~\bibnamefont {Chen}}, \bibinfo
  {author} {\bibfnamefont {J.}~\bibnamefont {Jia}}, \bibinfo {author}
  {\bibfnamefont {X.}~\bibnamefont {Dai}}, \bibinfo {author} {\bibfnamefont
  {Z.}~\bibnamefont {Fang}}, \bibinfo {author} {\bibfnamefont {S.-C.}\
  \bibnamefont {Zhang}}, \bibinfo {author} {\bibfnamefont {K.}~\bibnamefont
  {He}}, \bibinfo {author} {\bibfnamefont {Y.}~\bibnamefont {Wang}}, \bibinfo
  {author} {\bibfnamefont {L.}~\bibnamefont {Lu}}, \bibinfo {author}
  {\bibfnamefont {X.-C.}\ \bibnamefont {Ma}}, \ and\ \bibinfo {author}
  {\bibfnamefont {Q.-K.}\ \bibnamefont {Xue}},\ }\href {\doibase
  10.1126/science.1234414} {\bibfield  {journal} {\bibinfo  {journal}
  {Science}\ }\textbf {\bibinfo {volume} {340}},\ \bibinfo {pages} {167}
  (\bibinfo {year} {2013})},\ \Eprint
  {http://arxiv.org/abs/http://www.sciencemag.org/content/340/6129/167.full.pdf}
  {http://www.sciencemag.org/content/340/6129/167.full.pdf} \BibitemShut
  {NoStop}%
\bibitem [{\citenamefont {Xu}\ \emph {et~al.}(2014)\citenamefont {Xu},
  \citenamefont {Miotkowski}, \citenamefont {Liu}, \citenamefont {Tian},
  \citenamefont {Nam}, \citenamefont {Alidoust}, \citenamefont {Hu},
  \citenamefont {Shih}, \citenamefont {Hasan},\ and\ \citenamefont
  {Chen}}]{xu14}%
  \BibitemOpen
  \bibfield  {author} {\bibinfo {author} {\bibfnamefont {Y.}~\bibnamefont
  {Xu}}, \bibinfo {author} {\bibfnamefont {I.}~\bibnamefont {Miotkowski}},
  \bibinfo {author} {\bibfnamefont {C.}~\bibnamefont {Liu}}, \bibinfo {author}
  {\bibfnamefont {J.}~\bibnamefont {Tian}}, \bibinfo {author} {\bibfnamefont
  {H.}~\bibnamefont {Nam}}, \bibinfo {author} {\bibfnamefont {N.}~\bibnamefont
  {Alidoust}}, \bibinfo {author} {\bibfnamefont {J.}~\bibnamefont {Hu}},
  \bibinfo {author} {\bibfnamefont {C.-K.}\ \bibnamefont {Shih}}, \bibinfo
  {author} {\bibfnamefont {M.~Z.}\ \bibnamefont {Hasan}}, \ and\ \bibinfo
  {author} {\bibfnamefont {Y.~P.}\ \bibnamefont {Chen}},\ }\href
  {http://dx.doi.org/10.1038/nphys3140} {\bibfield  {journal} {\bibinfo
  {journal} {Nat Phys}\ }\textbf {\bibinfo {volume} {10}},\ \bibinfo {pages}
  {956} (\bibinfo {year} {2014})}\BibitemShut {NoStop}%
\end{thebibliography}%


\begin{figure*}[!t]
 \includegraphics[width=8cm]{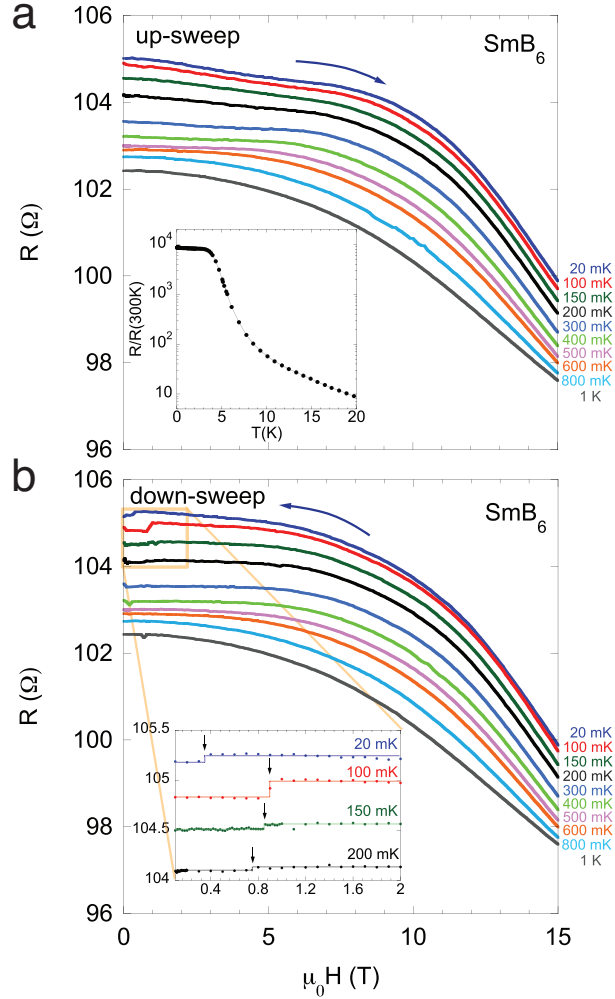}
 \caption{{\bf Low-temperature magnetoresistance of topological Kondo insulator \smb.} 
 {\bf a}, Magnetoresistance measured in perpendicular magnetic field orientation $(H \parallel [001], I \parallel [100])$ upon increasing magnetic field, taken at constant temperatures between 20~mK and 1~K, where surface conduction dominates that of the bulk material. The temperature dependence of resistivity in zero field is shown in the inset, normalized to its room temperature value. 
  Upon decreasing temperature below 500~mK, there is an apparent oscillation in field.
{\bf b}, Magnetoresistance taken in the same perpendicular field orientation but measured upon down-sweep of field, showing qualitatively similar behavior as for up-sweep data, but with notable differences including a strong suppression of the oscillatory amplitude present in up-sweep data, as well as abrupt transition-like jumps in the data as highlighted in the inset zoom (solid lines are guides to the eye). These differences are ascribed to the presence of ferromagnetism, as described in the text. }
\end{figure*}

\begin{figure*}[t]
 \includegraphics[width=15cm]{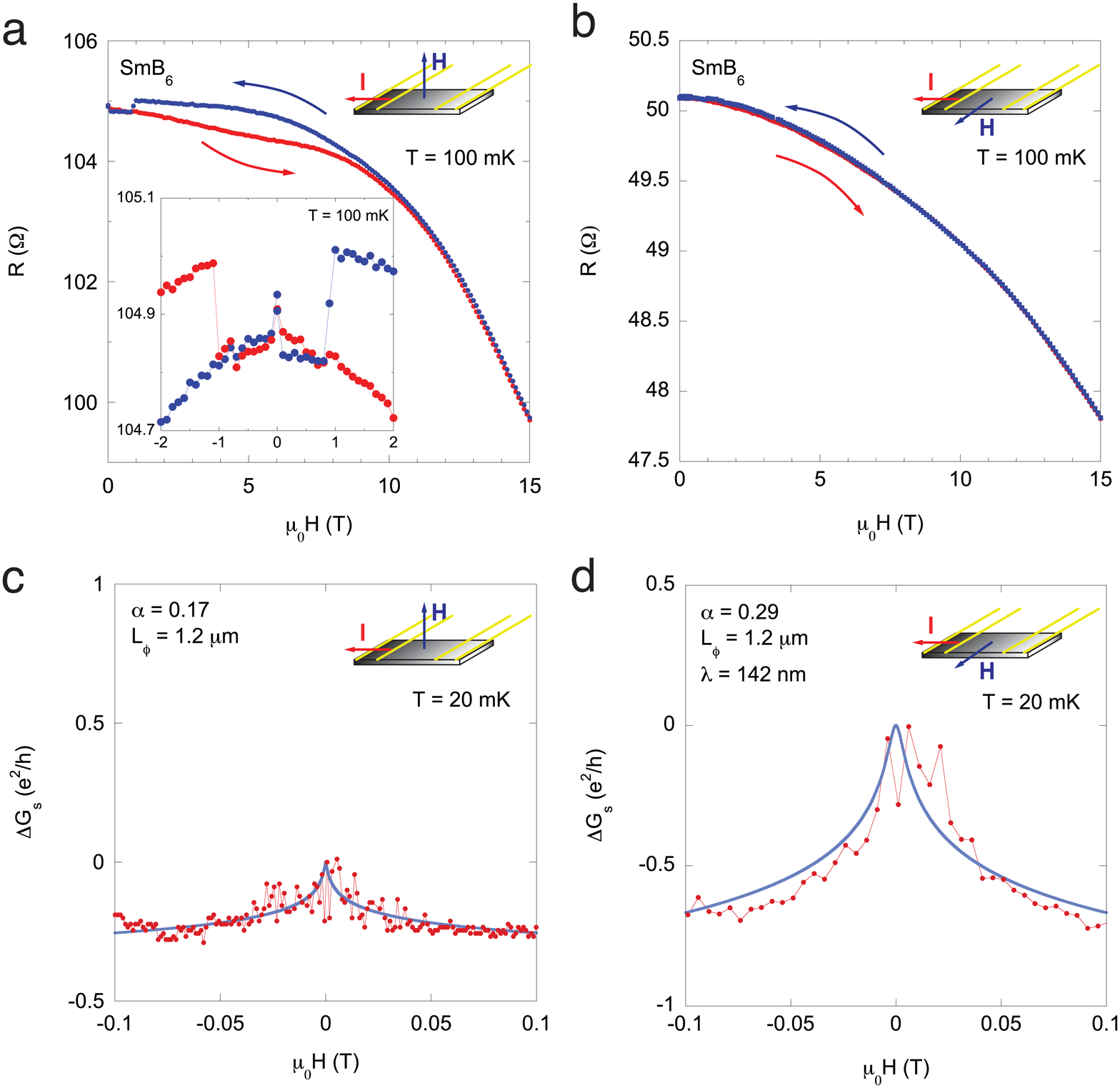}
 \caption{{\bf Anisotropy of low-temperature magnetoresistance hysteresis and weak antilocalization in \smb.}  
 {\bf a}, Comparison of magnetoresistance measured at 100~mK upon sweeping magnetic field up and down (see arrows) with field orientation perpendicular to the large surface of the sample with contacts. A hysteretic loop is evident between a turning field of $\sim 10$~T and a discontinuous jump at low field, and does not depend on field sign as highlighted by the inset zoom.   
 {\bf b}, The hysteresis is greatly suppressed when field is oriented parallel to the measurement plane, as shown for the same sample as in {\bf a} (with contacts reapplied).
 {\bf c}, Weak antilocalization is also observed at low fields and temperatures, but with surprisingly small coefficient of $\alpha$=0.17 (see text) for perpendicular field orientation. 
 {\bf d}, When field is applied parallel to the measurement plane, the weak antilocalization correction is much larger with $\alpha$=0.29 (see text), indicating a very strong anisotropy opposite to that expected for the usual orbital configuration originating from the finite penetration depth $\lambda$, but consistent with strong spin-flip scattering. Solid lines are fits to the data to WAL correction formula (see text). }
\end{figure*}

\begin{figure*}[p]
 \includegraphics[width=17cm]{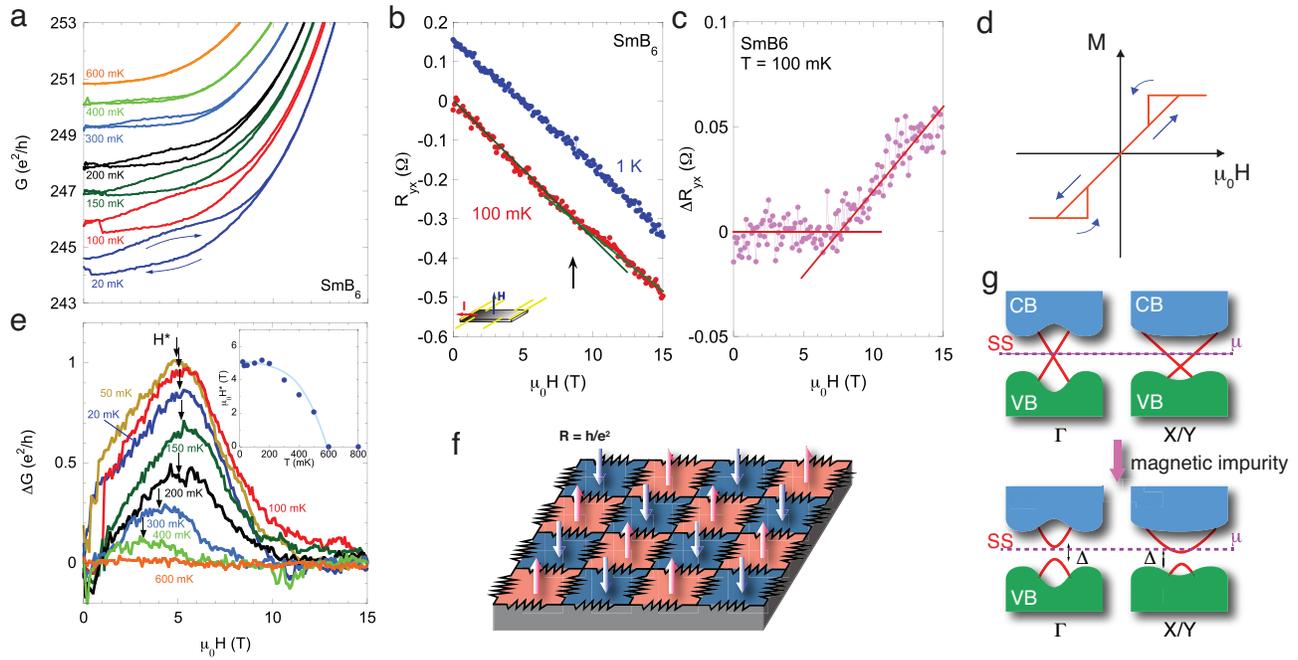}
 \caption{{\bf Quantum conductance along surface magnetic domain wall edges.} 
 {\bf a}, Magnetoconductance $G$ of {\smb} for perpendicular field orientation (20~mK and 100~mK data is vertically offset by -1.2$e^2/h$ and -0.3$e^2/h$, respectively, for clarity). The closure of the hysteretic loop with increasing temperature is consistent with the loss of surface ferromagnetism at a Curie temperature of $\sim 600$~mK. The enhanced (decreased) conductance upon up(down)-sweep is opposite to that expected in conventional ferromagnetic metals, providing an indication of the topological nature of the surface conducting states (see text). 
{\bf b}, Anomalous Hall effect in the low-temperature Hall resistance for sample \#1. A prominent kink is present in low-temperature (100~mK) data near the turning field of 8~T at which the hysteresis loop closes, indicating the presence of ferromagnetic order. The kink disappears at higher temperatures (1~K), where the hysteresis in magnetoresistance also vanishes. 
{\bf c}, Field dependence of the difference of Hall resistance $\Delta R_{yx} = R_{yx} - AH$, where $A$ is a linear coefficient obtained from a fitting below 5~T at 100~mK. An obvious onset associated with the anomalous Hall effect is observed at 8~T.
{\bf d}, Schematic hysteresis curve in magnetization of an exchange bias system.
{\bf e}, Magnetic field dependence of the difference of up- and down-sweep magnetoconductance $\Delta G$=$G_{up}$--$G_{dn}$. The appearance of one quantum of conductance $e^2/h$ in the low temperature limit provides evidence for a scenario where surface ferromagnetism gaps the topological conducting states within magnetic domains on the surface, limiting conductance to the domain walls. Arrows indicate the characteristic field $H^{\ast}$ where a maximum in $\Delta G$ is associated with a coercive field (see text). The inset presents the evolution of $H^{\ast}$, consistent with the onset of ferromagnetic order at 600~mK.
{\bf f}, schematic representation of the proposed grid-like ferromagnetic domain structure with walls separating domains of oppositely oriented moments, forming a virtual infinite resistance network with quantized components $R=h/e^2$. The resistance between any two non-adjacent nodes in such a network is on the order of $R$.
{\bf g}, Schematic band structure for {\smb} at the $\Gamma$ and $X/Y$ high-symmetry points. Due to the localized ordered moments, energy gaps are induced at the Dirac points, meaning the Dirac bands become massive. Because the chemical potential $\mu$ is fixed at one value, it likely falls within the gap of one band but not in the other (see text).}
\end{figure*}

\end{document}